\documentclass[epjCONF]{svjour}

\usepackage{graphicx,bm}
\usepackage{amsmath,amssymb}
\usepackage{mathrsfs,verbatim,subfigure}
\usepackage{graphics}
\usepackage[varg]{txfonts} 
\usepackage[latin1]{inputenc}

\session-title{Resonance Workshop at UT Austin}

\begin{document}

\title{Evaluating chiral symmetry restoration through the use of sum rules}

\author{Paul~M.~Hohler \thanks{\email{pmhohler@comp.tamu.edu}} \and Ralf~Rapp \thanks{\email{rapp@comp.tamu.edu}}}

\institute{Cyclotron Institute and Department of Physics and Astronomy, Texas A\&M University, College Station, TX 77843-3366, USA}

\abstract{
We pursue the idea of assessing chiral restoration via in-medium
modifications of hadronic spectral functions of chiral partners.
The usefulness of sum rules in this endeavor is illustrated, focusing
on the vector/axial-vector channel.
We first present an update on constructing quantitative results for
pertinent vacuum spectral functions. These spectral functions serve
as a basis upon which the in-medium spectral functions can be constructed.
A striking feature of our analysis of the vacuum spectral functions
is the need to include excited resonances, dictated by satisfying
the Weinberg-type sum rules. This includes
excited states in {\emph both} the vector and axial-vector channels.
Preliminary results for the finite temperature vector spectral
function are presented. Based on a $\rho$ spectral function tested
in dilepton data which develops a shoulder at low energies, we find that
the $\rho^\prime$ peak flattens off.
The flattening may be a sign of chiral restoration, though a study of
the finite temperature axial-vector spectral function remains to be
carried out.}

\maketitle

\section{Introduction} \label{sec:intro}

It is a long-standing problem in nuclear physics to observe
chiral symmetry restoration in QCD matter. Lattice QCD computations of
order parameters of the chiral transition, most notably the quark condensate,
find the latter to vanish at high temperature indicative of chiral
restoration~\cite{Borsanyi:2010bp,Bazavov:2011nk}. Unfortunately, the quark
condensate cannot be directly measured in experiment. Therefore,
one must rely on other more indirect methods to observe chiral symmetry
restoration.

A characteristic property of hadronic resonances are their spectral
functions. In vacuum, resonances give rise to peak-like structures
which are usually well described by a Breit-Wigner form. In medium,
the hadronic states interact with the heat-bath particles inducing
substantial changes of the peak-like structures in the spectral
functions~\cite{Friman:2011zz}.
Since chiral restoration is a consequence of interactions with the medium,
by studying the medium modifications of the spectral functions, one
hopes to deduce signatures of chiral restoration. Yet, this raises several
questions, for example: What type of medium modifications are expected from
chiral symmetry restoration? Do different hadronic channels share generic
features when approaching chiral restoration, or must we
look at a large number of different resonances to gleam chiral restoration?
Sum rules provide some of the answers to these questions.

Sum rules, in general, relate the spectral functions via a
dispersion relation to low-energy condensates, including the quark
condensate among others. The spectral medium modifications expected at
finite temperature are then encoded in the sum rules
by changes in the condensates. Therefore, if one knows the in-medium
spectral functions, from either experiment or theory, one can make
predictions on the properties of the condensates, or visa versa. We
will focus on two classes of sum rules here, QCD sum rules and
Weinberg-type sum rules.

QCD sum rules were first introduced in
Refs.~\cite{Shifman:1978bx,Shifman:1978by} by relating the spectral
function in a particular hadronic channel to an operator product expansion
(OPE) containing all operators relevant for that channel (usually
truncated in a series of inverse momentum transfer). The
QCD sum rule is the most general relation of a spectral function to the
OPE for a single hadronic channel, including both
chirally symmetric and chirally breaking operators. Therefore, the medium
modifications of the spectral function are driven by both types of operators.
This introduces the problem that if one observes medium modifications of
the spectral function, one cannot attribute these solely to chiral
restoration. Thus, chiral restoration cannot be deduced by examining
medium modifications of a spectral function in a single channel.
However, if one considers hadronic states which are chiral partners,
then the effects of the chirally symmetric operators can be canceled out so
that only the chirally breaking operators survive.
Therefore, to study and observe chiral symmetry restoration one is
compelled to study the medium modifications of chiral partners.

The appropriate set of sum rules to study chiral partners are Weinberg-type
sum rules. They refer to a set of sum rules which relate the difference
of spectral function of two chiral partners to chirally breaking
operators. They were developed by Weinberg~\cite{Weinberg:1967kj} using
current algebra relating the isovector vector and axial-vector channels.
Subsequently, other sum rules of similar form have been
considered~\cite{Das:1967ek,Kapusta:1993hq}, including extensions to
chiral partners other than vector and axial-vector light
mesons~\cite{Hilger:2011cq}. The Weinberg-type sum rules have
also been extended into the medium~\cite{Kapusta:1993hq}. Thereby,
the medium modification of the spectral functions of the chiral
partners can be related to chiral-breaking operators.

The pair of chiral partners that has been studied the most are the isovector
vector and axial-vector light mesons, {\it i.e.}, the $\rho$- and $a_1$-meson
channels. The spectral functions for both channels have been accurately
 measured below $s=3 {\rm GeV}^2$ by ALEPH~\cite{Barate:1998uf}
and OPAL~\cite{Ackerstaff:1998yj} via $\tau$ decays into an even and odd
number of pions. Dilepton data from heavy-ion collisions
provide information about the in-medium $\rho$ spectral
function~\cite{Arnaldi:2006jq,Adamova:2006nu}, while the in-medium
axial-vector spectral function has not been experimentally studied.
Until that time, we are left to construct models of the
axial-vector spectral function to study its medium modifications.
Sum rules can provide an important guide in these constructions. In
such a set-up, rather than inferring the condensates from the spectral
functions, one can use the temperature dependence of the condensates,
as computed, {\it e.g.}, in thermal lattice QCD, as an input to determine
the spectral functions. In order to accurately constrain the features
of the spectral function which are similar and those which are dissimilar
between the vector and axial-vector channels, {\emph both} QCD and
Weinberg-type sum rules should be considered.

Several aspects of the idea of using sum rules and the known behavior of condensates in medium
to ascertain properties of the in-medium spectral functions has thus
far been addressed (see, {\it e.g.}, Part II, Sec.~2 of
Ref.~\cite{Friman:2011zz} for a review), both at finite density or at
finite temperature; we here focus on the latter. Among the differences
between the previous studies are the ans\"{a}tze adopted for the spectral
functions. While the high-energy part of the spectral functions has
been routinely implemented by a continuum with a sharp threshold onset,
the modeling of the low-energy resonance part has significantly evolved.
Early works used a delta function to represent the ground-state resonance,
while later on more realistic spectral Breit-Wigner spectral functions
have been employed. In the medium, most of the previous works used QCD
sum rules while few evaluated the Weinberg-type sum rules.
Despite the wide range of ans\"{a}tze, a few general trends have been
established. The strength of the ground-state resonance, either the
$\rho$ or $a_1$, tends to lower energies, either through a decreasing
peak position or increasing peak width (when included), or both.
Furthermore, the energy associated with the threshold of the continuum
also tends to shift to lower energies. In some studies,
these trends were more pronounced than in others.

We intend to examine chiral symmetry restoration by ultimately studying
medium modifications of spectral functions for the vector and axial-vector
channels using experimental data (when available) combined withe QCD and
Weinberg-type sum rules. The starting point must be a realistic description in
the vacuum, where spectral functions can be quantitatively constrained
using the $\tau$-decay data from ALEPH and further tested by Weinberg-type
sum rules. The constructed spectral functions can then be used along with
the QCD sum rules to determine viable vacuum values of the condensates. These
results were presented in Ref.~\cite{Hohler:2012xd} and will be summarized
below. These vacuum spectral functions serve as a basis for studying medium
modifications at finite temperature. This investigation first addresses
the construction of finite-temperature spectral functions in the vector channel
using the QCD sum rules as a constraint. Future work will be devoted to the
construction of the axial-vector spectral function at finite temperature such
that they satisfy both the QCD and the Weinberg-type sum rules.

This paper is organized as follows. Section~\ref{sec:sum} summarizes sum
rules in vacuum and at finite temperature. In particular, it illustrates
their role in studying chiral symmetry restoration.
This is followed by Sec.~\ref{sec:spec} which describes the results for the
constructed spectral functions in vacuum and for the vector channel at
finite temperature. Conclusions are given in Sec.~\ref{sec:con}.

\section{Sum Rules}
\label{sec:sum}
\subsection{Vacuum}
\label{sec:sumVac}
Let us consider the current-current correlator in the vector channel, defined as
\begin{equation}
\Pi^{\mu\nu}_V(q) = - i \int d^4 x e^{i q x} \langle T j_V^\mu(x) j_V^\nu(0)\rangle.
\end{equation}
In vacuum, the correlator can be expressed in terms of a single function,
the polarization function, $\Pi(q^2)$. The QCD sum rule amounts to the
statement that this polarization function can be calculated in two different
ways: either by a dispersion relation, or through an operator product
expansion (OPE) for large values of $Q^2=-q^2>0$. After a so-called
Borel transform it reads
\begin{equation}
\label{eq:rhoQCD}
\frac{1}{M^2}\!\int_0^\infty \!ds \frac{\rho_V(s)}{s} e^{-s/M^2}
= \frac{1}{8\pi^2} \left(1+\frac{\alpha_s}{\pi}\right)
+\frac{m_q \langle\bar{q}q\rangle}{M^4}
+\frac{1}{24 M^4}\langle\frac{\alpha_s}{\pi} G_{\mu\nu}^2\rangle
- \frac{56 \pi \alpha_s}{81 M^6}  \langle \mathcal{O}_4^V \rangle \ldots \, ,
\end{equation}
where $\langle\bar{q}q\rangle$ is the quark condensate,
$\langle\frac{\alpha_s}{\pi} G_{\mu\nu}^2\rangle$ the gluon condensate,
$\langle \mathcal{O}_4^V \rangle$ the vector 4-quark condensate (explicitly
given, {\it e.g.}, by Eq.~(2.19) in Ref.~\cite{Leupold:2001hj}) and
$\rho_V\equiv -{\rm Im} \Pi_V/\pi$ the vector spectral function. Typically
one assumes that the 4-quark condensate can be factorized such that
$\langle \mathcal{O}_4^V \rangle = \kappa_V \langle \bar{q}q \rangle^2$,
where $\kappa_V$ is a parameter accounting for deviations from
``factorization". A similar expression can be derived for the axial-vector
channel~\cite{Shifman:1978bx,Shifman:1978by}
\begin{equation}
\label{eq:a1QCD}
\frac{1}{M^2}\!\int_0^\infty \!ds \frac{\bar{\rho}_A(s)}{s} e^{-s/M^2}
= \frac{1}{8\pi^2} \left(1+\frac{\alpha_s}{\pi}\right)
+\frac{m_q \langle\bar{q}q\rangle}{M^4}
+\frac{1}{24 M^4}\langle\frac{\alpha_s}{\pi} G_{\mu\nu}^2\rangle
+ \frac{88 \pi \alpha_s}{81 M^6}  \langle \mathcal{O}_4^A \rangle \ldots \, ,
\end{equation}
where $\bar{\rho}_A$ is the axial-vector spectral function including
the pion pole. As in the vector channel, the
axial-vector spectral function is related to the quark and gluon
condensates, but a different 4-quark condensate (see, {\it e.g.},
Eq.~(2.20) in Ref.~\cite{Leupold:2001hj} for its explicit form).
One may also factorize this operator as
$\langle \mathcal{O}_4^A \rangle = \kappa_A \langle \bar{q}q \rangle^2$.
In principle, $\kappa_V$ and $\kappa_A$ are numerically distinct parameters,
however, for the results presented here, we have assumed that they take
on the same value. Note that the difference between the OPE for $\rho_V$ and
$\bar{\rho}_A$ in vacuum is entirely given by the 4-quark condensate term.
As mentioned above, the spectral functions are related to both chirally
symmetric and chirally breaking operators. Therefore one can not deduce
chiral restoration from either one channel separately.

Let us consider the difference between the vector and axial-vector QCD sum
rules. On the left-hand-side (LHS), one simply has the difference of the
two spectral functions, while on the right-hand-side (RHS) everything except
the 4-quark condensates cancels. The difference in the spectral functions
should be well-enough behaved so that the Borel exponential can be Taylor
expanded.
By equating the coefficients of same powers of $1/M^2$ on each side, a new
set of sum rules can be constructed, namely
\begin{eqnarray}
&({\rm WSR}\, 1)& \quad \quad \int_0^\infty \!ds \,
\frac{\Delta\rho(s)}{s} = f_\pi^2 \ ,
\label{eq:WSR1} \\
&({\rm WSR}\, 2)& \quad \quad \int_0^\infty\! ds\, \Delta\rho(s)
=   f_\pi^2 m_\pi^2 = -2 m_q \langle \bar{q}q \rangle \ ,
\label{eq:WSR2}\\
\label{eq:WSR3}
&({\rm WSR}\, 3)& \quad \quad \int_0^\infty ds\, s \,\Delta\rho(s)
= - 2 \pi \alpha_s \langle \mathcal{O}_4 \rangle,
\end{eqnarray}
where $\Delta\rho \equiv \rho_V - \rho_A$. The contribution from the
pion pole has been moved the RHS, and a new chirally breaking 4-quark
condensate has been introduced,
\begin{equation}
\langle \mathcal{O}_4 \rangle = \frac{16}{9} \left( \frac{7}{18}\langle \mathcal{O}_4^V \rangle + \frac{11}{18} \langle \mathcal{O}_4^A \rangle\right).
\end{equation}
Factorization can also be applied to this 4-quark condensate,
$\langle \mathcal{O}_4\rangle = 16/9 \kappa \langle \bar{q}q \rangle^2$
with $\kappa = 7/18 \kappa_V + 11/18 \kappa_A$.
The first two equations are the Weinberg sum rules \cite{Weinberg:1967kj},
with a finite quark-mass correction in the second equation \cite{Pascual:1981jr,Narison:1981ra,Peccei:1986hw,Dmitrasinovic:2000ei},
while the third equation is derived in \cite{Kapusta:1993hq}.
Furthermore, a sum rule derived in \cite{Das:1967ek} is associated with
Weinberg-type sum rules though it is not directly derived from the QCD
sum rules as the others. It reads
\begin{equation}
\label{eq:WSR0}
({\rm WSR}\, 0) \quad \quad \int_0^\infty \!ds\,
\frac{\Delta\rho(s)}{s^2} =
\frac{1}{3} f_\pi^2 \langle r_\pi^2\rangle -F_A \ ,
\end{equation}
where $\langle r_\pi^2\rangle$ is the mean squared radius of the charged
pion and $F_A$ the coupling constant for the radiative pion decay.

This simple derivation illustrates the importance of considering chiral
partners.
The OPE of the QCD sum rule is replaced by only chirally breaking operators.
An ``arbitrary" collection of hadronic-resonance QCD sum rules would not have
resulted in such simplifications. The Weinberg-type sum rule also show
that by knowing the vector and axial-vector spectral functions, one can
infer the quark condensate along with additional chiral order parameters.

\subsection{Finite Temperature}
\label{sec:sumTemp}
The finite-temperature extensions of both QCD and the Weinberg-type
sum rules have been considered in \cite{Hatsuda:1992bv} and
\cite{Kapusta:1993hq}, respectively. We will summarize and highlight some
key points. At finite temperature, Lorentz symmetry is broken and the
polarization function develops two components, usually characterized
by $\Pi_T$ and $\Pi_L$ for the 3D transverse and longitudinal
polarizations. Here we will focus on the case of vanishing 3-momentum,
$\vec{q}=0$, in which $\Pi_L=\Pi_T$ and we can drop the labels.

At finite temperature, the condensate-values change which the sum rules
translate into medium modifications of the spectral functions.
Obviously, the Weinberg-type sum rules are sensitive to only chiral
breaking operators, while the QCD sum rules are sensitive to both
chirally breaking and chirally symmetric. The QCD sum rules are particularly
useful when the focus is on a single channel, not the relation between
different channels.

It turns out that the construction of the finite-temperature Weinberg sum
rules simply amounts to replacing the vacuum spectral functions and
operators with their values at finite temperature in the expressions
above~\cite{Kapusta:1993hq}. The pion pole in general develops a non-trivial
spectral distribution  and so its contribution is included back into the
axial-vector spectral function.  One has
\begin{eqnarray}
&({\rm WSRFT}\, 1 )& \quad \quad \int_0^\infty \!ds \,
\frac{\Delta\bar{\rho}(s)}{s} =0 \ ,
 \\
&({\rm WSRFT}\, 2)& \quad \quad \int_0^\infty\! ds\, \Delta\bar{\rho}(s)
=  0 ,
\\
&({\rm WSRFT}\, 3)& \quad \quad \int_0^\infty ds\, s \,\Delta\bar{\rho}(s)
= - 2 \pi \alpha_s \langle \mathcal{O}_4 \rangle_T,
\end{eqnarray}
where $\Delta\bar{\rho}(s) = \rho_V(s) - \bar{\rho_A}(s)$. A derivation
from the finite temperature QCD sum rules is also possible.
These equations show that at chiral restoration where the pion pole and
the 4-quark condensate vanish, the vector and
axial-vector spectral functions should be degenerate.

For the QCD sum rules, the vacuum spectral function is replaced by the finite
temperature spectral function, while the OPE becomes more involved. In
addition to the scalar condensates, which figure in vacuum and develop a
temperature dependence, new non-scalar condensates appear.
The latter are possible because of the breaking of Lorentz symmetry in
medium. They are classified by their dimension, spin, and twist,
where twist = dimension - spin (scalar operators have twist-0). The resulting
sum rule reads
\begin{equation}
\begin{split}
\label{eq:QCDT}
\frac{1}{M^2}\!\int_0^\infty \!ds \frac{\rho_V^T(s)}{s} e^{-s/M^2}
= &\frac{1}{8\pi^2} \left(1+\frac{\alpha_s}{\pi}\right)
+\frac{m_q \langle\bar{q}q\rangle_T}{M^4}
+\frac{1}{24 M^4}\langle\frac{\alpha_s}{\pi} G_{\mu\nu}^2\rangle_T
- \frac{56 \pi \alpha_s}{81 M^6}  \langle \mathcal{O}_4^V \rangle_T \\&+ \frac{\langle \mathcal{O}^{d=4, \tau=2} \rangle_T}{M^4}+\frac{\langle \mathcal{O}^{d=6, \tau=2} \rangle_T}{M^6}+\frac{\langle \mathcal{O}^{d=6, \tau=4} \rangle_T}{M^6} \, ,
\end{split}
\end{equation}
where $\langle \mathcal{O}^{d=4, \tau=2} \rangle_T$,
$\langle \mathcal{O}^{d=6, \tau=2} \rangle_T$, and $\langle \mathcal{O}^{d=6, \tau=4} \rangle_T$
represent the twist-2 and twist-4 operators. A discussion of the higher-twist
operators can be found in \cite{Hatsuda:1992bv,Leupold:1998bt,Zschocke:2002mn}.
A similar expression can be derived for the axial-vector channel. Note that all
of the higher-twist operators are chirally symmetric and so are identical
between the two channels.

\begin{figure}[htb]
\centering
\subfigure[Quark condensate]{\label{fig:q2T}
\includegraphics[width=.45\textwidth]{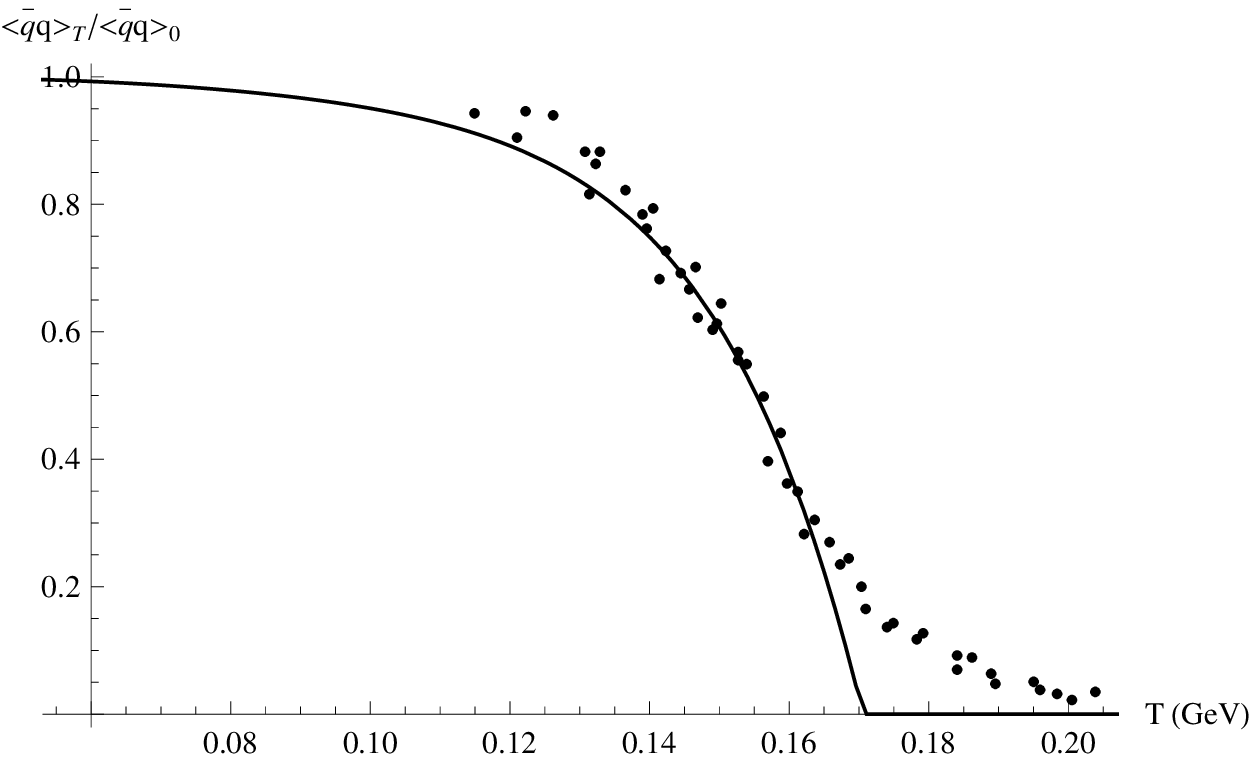}}
\subfigure[4-Quark condensate]{\label{fig:q4T}
\includegraphics[width=.45\textwidth]{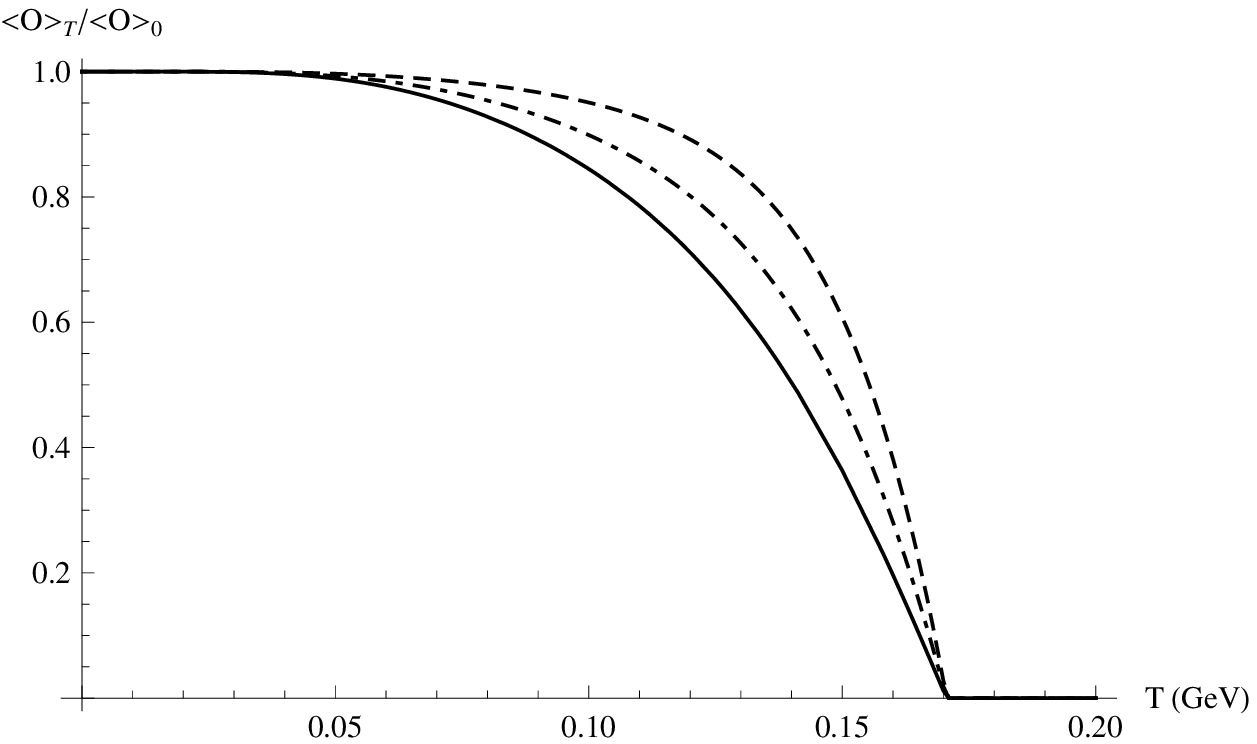}}
\caption{a) Temperature dependence of quark condensate relative to its vacuum value compared with lattice data points \cite{Borsanyi:2010bp}. b) Temperature dependence of vector (solid curve) and axial-vector (dot-dashed curve) 4-quark condensates relative to their vacuum values compared with the temperature dependence of the quark condensate relative to its vacuum value (dashed curve).\label{fig:condT} }
\end{figure}

To determine the modifications of the condensates at low temperatures,
one may invoke the dilute gas approximation which amounts to a generic
expression of type
\begin{equation} \label{eq:opT}
\langle \mathcal{O} \rangle_T \simeq \langle \mathcal{O}\rangle_0 + d_X \int \frac{d^3 k}{\left(2 \pi\right)^2 2 E_k} \langle X(\vec{k})|\mathcal{O}|X(\vec{k})\rangle,
\end{equation}
where $\langle \mathcal{O}\rangle_0$ is the vacuum value of a given operator
$\langle X(\vec{k})|\mathcal{O}|X(\vec{k})\rangle$ is the matrix element
of the same operator within hadron $X$, and $d_X$ is the degeneracy factor
for that hadron. For the vacuum values of the operators we follow
Ref.~\cite{Hohler:2012xd}, namely
$\langle \bar{q}q \rangle = (-0.25 {\rm GeV})^3$,
$\langle \frac{\alpha_s}{\pi}G^2_{\mu\nu}\rangle = 0.022 {\rm GeV}^4$,
$\langle \mathcal{O}_4^V \rangle = \kappa \langle \bar{q}q \rangle^2$ with
$\kappa = 2.1$. The expansion (\ref{eq:opT}) also holds for the higher-twist
operators with a vanishing vacuum value. This reduces the problem to
determining the
matrix elements $\langle X(\vec{k})|\mathcal{O}|X(\vec{k})\rangle$. To do so,
we assume that the medium can be described by a hadron resonance gas (HRG)
along the lines of \cite{Leupold:2006ih}. All reliably known hadrons with a mass
less than $2$ GeV \cite{pdg} have been considered. The temperature dependence
for the quark and 4-quark condensates is augmented around the transition temperature
beyond the consideration of the HRG such that the quark condensate better fits
lattice calculations and that the 4-quark condensates vanish at the same
temperature as the quark condensate. The results are depicted in
Fig.~\ref{fig:condT}. Further details concerning the HRG estimate of the
higher-twist operators will be presented in future work. The temperature
dependence of the condensates increases the OPE side relative to the vacuum.
Via the QCD sum rule, this increase is manifest in the spectral function
through an increased spectral strength at lower energies. This will be
apparent when we discuss the results for the finite-temperature vector
spectral function below.

\section{Spectral Functions and Numerical Evaluation}
\label{sec:spec}
Having described the relevant sum rules in vacuum and shown how they are
modified at finite temperature, we are now in position to construct model
spectral functions, including experimentally determined properties, and test
whether their medium modifications satisfy the sum rules. We shall
begin in the vacuum and construct spectral functions for both the vector and
axial-vector channels. The results presented here are a summary of those found
in \cite{Hohler:2012xd}.

For both channels, the spectral function is constructed from three parts,
\begin{equation}
\rho_V(q_0) = \rho_V^{\rm gs}(q_0) + \rho_V^{\rm ex}(q_0) + \rho^{\rm cont}(q_0)
\end{equation}
where $\rho_V^{\rm gs}$, $\rho_V^{\rm ex}$ and $\rho_V^{\rm cont}$ correspond to
the ground-state resonance, an excited-state resonance and a perturbative QCD (pQCD)
continuum, respectively. There are three critical features about this ansatz.
First, for the $\rho(770)$ contribution to the spectral function, we employ
the microscopic model of \cite{Urban:1999im}, which satisfies the
low-energy $\tau$-decay data well~\cite{Rapp:2002tw}. It has the added
advantage that its medium modifications have been computed and successfully
applied to experiment \cite{Rapp:1999us,Riek:2008ct}. This provides a
strong basis for forthcoming investigations in medium.
Second, the continuum contribution for the vector and the
axial-vector channels is postulated to be identical. This is to be expected
at higher energies where pQCD can be used to calculate its contribution.
We have extended this premise to all energies, and chosen a continuous
onset with energy, {\it i.e.} a discontinuous threshold is not considered.
Third, the first excited states are postulated (which, in fact, emerges as a
consequence of the degenerate continuum). The pertinent spectral
functions for the $\rho^\prime$ and $a_1^\prime$
are parameterized Breit-Wigner functions (as is the $a_1$, as a placeholder
for a forthcoming microscopic model).

\begin{figure}[htb]
\centering
\subfigure[Vector Spectral Function]{\label{fig:rhospec}
\includegraphics[width=.45\textwidth]{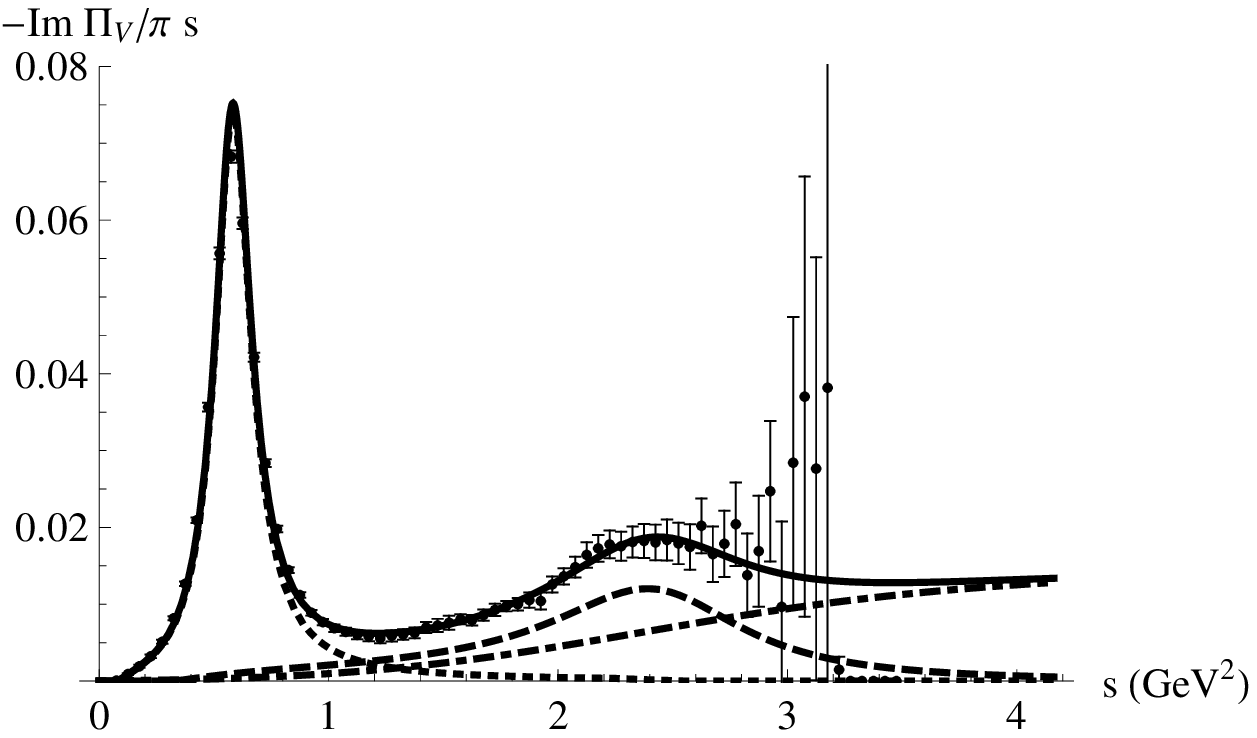}}
\subfigure[Axial-Vector Spectral Function]{\label{fig:a1spec}
\includegraphics[width=.45\textwidth]{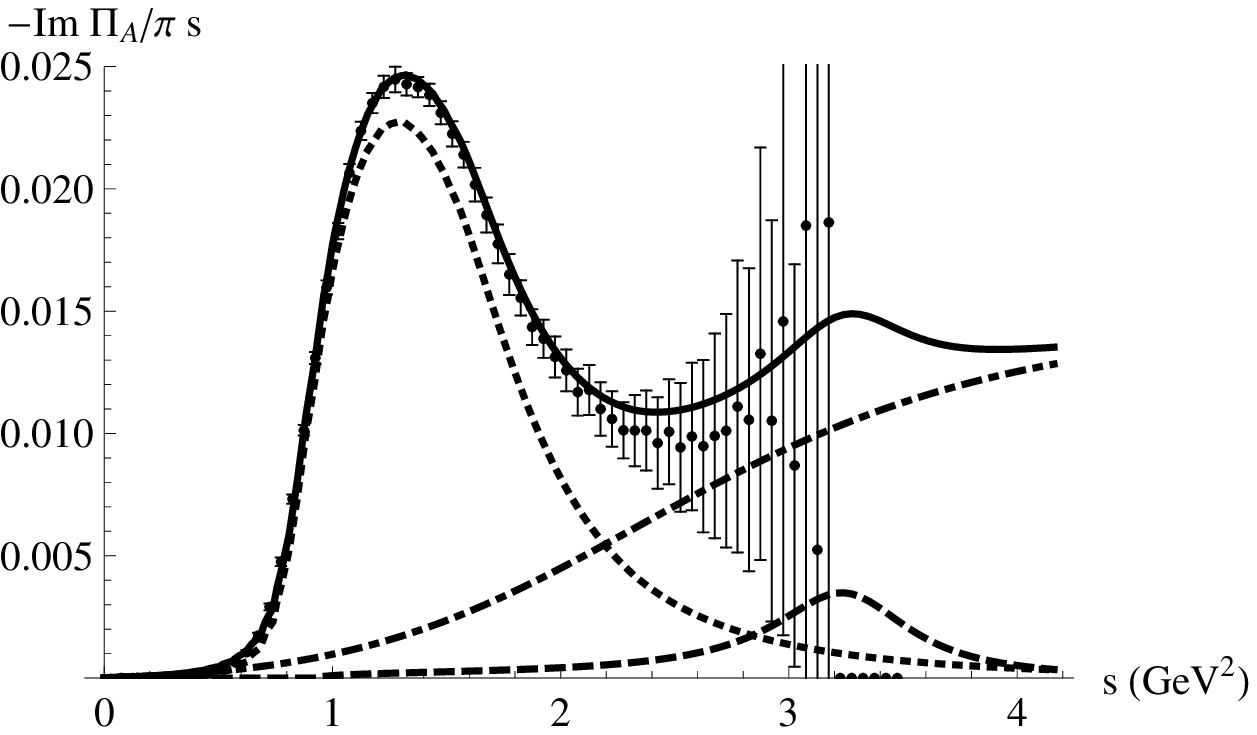}}
\caption{Spectral functions for the vector and axial-vector channels
compared to experimental data for hadronic $\tau$ decays by the ALEPH
collaboration~\cite{Barate:1998uf}. The different curves highlight the
contributions to the total spectral function (solid curve) from the
ground-state resonance (dotted curve), the excited resonance (dashed
curve), and the continuum (dot-dashed curve).}
\label{fig:vacspec}
\end{figure}

The parameters of the model were chosen such that the resulting spectral
functions satisfactorily agree with the $\tau$-decay data and that
Weinberg-type sum rules $0-2$ are satisfied. The comparison of the spectral
functions with data is illustrated in Fig.~\ref{fig:vacspec}. We find that
by postulating identical pQCD the continua between the two channels, one
is forced to consider their onset occurring at higher energies as compared
with previous studies. This is due to the relatively large valley in the
axial-vector spectral function around 2.2 ${\rm GeV}^2$. The larger
continuum onset, in turn, requires the inclusion of the $\rho^\prime$ so
that the vector spectral function agrees with the $\tau$-decay data, which
can be achieved with contributions from only the $\rho$, $\rho^\prime$ and
$a_1$ (plus continua). However, with this input the Weinberg-type sum rules
are violated, in a way that the LHS largely exceeds the RHS, the more so the
higher the moment. This clearly points to a missing ``high"-energy strength
in the axial-vector spectral function. We are therefore led to postulate the
presence of an excited axial-vector meson, namely the $a_1^\prime$.
There is large uncertainty in the $\tau$-decay data at higher energies,
sufficient enough to fit an excited state within the error bars (and/or
beyond the data range), as shown in Fig.~\ref{fig:a1spec}.
We deduce a mass of the excited state of around 1.8~GeV and a width of
around 200~MeV. In Table~\ref{tab:WSR}, we collect the resulting values
for the Weinberg-type sum rules
(a positive sign means that the vector channels contribution is greater).
The agreement with the three sum rules is much improved after inclusion
of the $a_1^\prime$, especially for WSR-0, -1 and -2, which we used as
criteria to infer the existence of the excited state.

The constructed spectral functions are subsequently used in the QCD sum
rules to determine the values of the gluon condensate and the factorization
parameters $\kappa_V$ and
$\kappa_A$. The vacuum values of these quantities are not as well established,
especially the factorization parameters. We assume that the two $\kappa$'s are
numerically the same. By minimizing the average deviation between the two
sides of the QCD sum rules as described
in \cite{Leinweber:1995fn,Leupold:1997dg,Hohler:2012xd}, we find
$\kappa = 2.1^{+.3}_{-.2}$ and
$\langle \frac{\alpha_s}{\pi} G^2_{\mu\nu} \rangle
= 0.022\pm 0.002\, {\rm GeV}^4$.
The optimized average deviations for vector and axial-vector channels
are 0.24\% and 0.56\%, respectively.

\begin{table}[htb]
\begin{center}
\begin{tabular}{|c|c|c|c|c|}
    \hline
    WSR &$0^{\rm th}$ & $1^{\rm st}$ & $2^{\rm nd}$ & $3^{\rm rd}$ \\
    \hline
    $\%$ agreement &-1.28\% & $\sim 0\%$ &$\sim 0\%$ & -96\% \\
    \hline
\end{tabular}
\end{center}
\caption{Percent disagreement between LHS and the RHS of the Weinberg-type
sum rules resulting from our fit.}
\label{tab:WSR}
\end{table}

The vacuum spectral functions can be used as a basis for investigating
medium modifications at finite temperature. For the remainder of this section,
preliminary results for the finite temperature vector spectral function will
be presented. The finite-temperature axial-vector spectral function will be
discussed in future work.

For the finite-temperature vector spectral function, we will primarily be
using the QCD sum rule to constrain it. We have argued above that the OPE
side will increase with temperature. This implies that the medium
modification of the spectral function will be such to increase its spectral
strength. In connection with the Borel transform, this is realized by
additional spectral strength at lower energies.

\begin{figure}[htb]
\centering
\includegraphics[width=.6\textwidth]{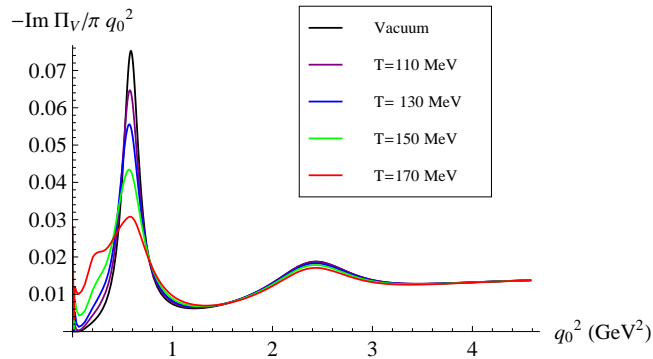}
\caption{(Color online) Spectral functions for the vector channel at different temperatures.}
\label{fig:vecsf}
\end{figure}

The ansatz for the finite-temperature spectral function is divided into the
same basic three parts as the vacuum ansatz: contributions from the ground
state, an excited state and from the pQCD continuum. As discussed
above, for the contribution from the $\rho$ meson, we use the spectral
function of \cite{Urban:1999im,Rapp:1999us}.
For the continuum, we postulate no temperature dependence. This is contrary
to previous studies which deduced a decrease in the continuum. However, as
we will demonstrate, we will still recover a spectral function which has
the appearance of a moving continuum, even if the continuum actually has
no temperature dependence. This assumption is also consistent with expectations
from pQCD. For the $\rho^\prime$, the medium modifications are implemented by
parameterizing the Breit-Wigner spectral function with adjustable mass,
width and coupling to the vector current. At each temperature, the
necessary adjustments to these parameters are made so that the QCD sum rule
is satisfied. As in the case of the vacuum, we use the average deviation as a
measure for the agreement of the QCD sum rules. We further insist that
parameters are monotonic with temperature; this ensures that un-physical
parameterizations such as a width narrowing with temperature do not occur.
The resulting spectral
functions at select temperatures are shown in Fig.~\ref{fig:vecsf}.

Two striking features of the finite-temperature spectral function are the
development of a low-energy shoulder to, and the reduction of, the $\rho$
peak (both pivotal to the description of dilepton data), as well as the
reduction of the $\rho^\prime$ peak. The formation of new low-energy
strength in the medium was to be expected from the QCD sum rules, however,
the shoulder actually adds too much strength.  Therefore the $\rho^\prime$
must reduce in strength (as shown) to compensate for
the extra strength in the $\rho$ peak. The reduction of the $\rho^\prime$ peak
(along with a slight broadening) flattens the higher energy region of the
spectral function. The resulting flattened spectral function starts to
resemble a flat continuum with a lower threshold. Therefore the effect which
has been previously associated with a moving continuum is in this model
realized via the dissociation of the $\rho^\prime$. That this dissociation
occurs prior to that of the $\rho$ is in line with the idea of sequential
dissociation of resonances with increasing temperature (not unlike heavier
quarkonia).
Lastly, the flattening of the spectral function may also signal chiral
restoration. The vacuum axial-vector spectral function has a majority of its
strength centered at an energy around the valley of the vector spectral
function. Thus, one could surmise that the filling of the "vector valley"
will be accompanied by a reduction of strength in the axial-vector channel
so that chiral restoration emerges, not unlike the well-known ``chiral
mixing" in a dilute pion gas~\cite{Dey:1990ba}. For a HRG this largely
remains a speculation until the medium modifications of the axial-vector
spectral function are determined more explicitly.

\section{Conclusion}
\label{sec:con}
We have illustrated the role of sum rules in examining chiral symmetry
restoration in hot matter. Results for the construction of vacuum spectral
functions for the vector and axial-vector channels have been presented.
Based on a quantitative analysis of existing vacuum data, we are led to
postulate the existence of an excited axial-vector resonance in order
to satisfy the Weinberg-type sum rules. Preliminary results from
in-medium vector spectral functions have also been presented. We have
constructed the latter starting from an in-medium $\rho$ spectral function
from microscopic hadronic effective field theory which is consistent with
SPS dilepton data. The spectral function further includes medium
modifications of the $\rho'$ peak as constrained by finite-temperature QCD
sum rules, while no medium modification to the continuum was considered.
We have found the resulting spectral functions to be consistent with the
idea that the medium broadens the states and reduces the strength of the
resonances, {\it i.e.}, both $\rho$ and $\rho'$. We have observed indications
for a sequential dissociation, and a rearrangement of spectral strength
toward the energy regions of the axial-vector resonance, suggestive of
chiral restoration.  However, it is important to emphasize that in order
to establish chiral restoration, either experimentally or theoretically,
quantitative knowledge of the spectral functions for both chiral partners
are required.  Therefore, the present study is only a beginning as the
medium modifications of the axial-vector spectral function remain
to be calculated. This work is currently underway.
Ideally this would also trigger a renewed emphasis to experimentally
measure the in-medium axial-vector spectral function, even in the
dilute phases of nuclear collisions or in elementary production
processes off ground-state nuclei.

\end{document}